# Common bibliometric approaches fail to assess correctly the number of important scientific advances for most countries and institutions

Alonso Rodríguez-Navarro[1,2] and Ricardo Brito[2]

[1] Departamento de Biotecnología - Biología Vegetal, Universidad Politécnica de Madrid, Avenida de Puerta de Hierro 2, 28040, Madrid, Spain

[2] Departamento de Estructura de la Materia, Física Térmica y Electrónica I GISC, Universidad Complutense de Madrid, Plaza de las Ciencias 3, 28040 Madrid, Spain

**Abstract**

Although not explicitly declared, most research rankings of countries and institutions are supposed to reveal their contribution to the advancement of knowledge. However, such advances are based on very highly cited publications with very low frequency, which can only very exceptionally be counted with statistical reliability. Percentile indicators enable calculations of the probability or frequency of such rare publications using counts of much more frequent publications; the general rule is that rankings based on the number of top 10% or 1% cited publications ($P_{top\ 10\%}$, $P_{top\ 1\%}$) will also be valid for the rare publications that push the boundaries of knowledge. Japan and its universities are exceptions, as their frequent Nobel Prizes contradicts their low $P_{top\ 10\%}$ and $P_{top\ 1\%}$. We explain that this occurs because, in single research fields, the singularity of percentile indicators holds only for research groups that are homogeneous in their aims and efficiency. Correct calculations for ranking countries and institutions should add the results of their homogeneous groups, instead of considering all publications as a single set. Although based on Japan, our findings have a general character. Common predictions of scientific advances based on $P_{top\ 10\%}$ might be severalfold lower than correct calculations.

# 1. INTRODUCTION

Technologically advanced countries invest large amounts of funds in research. Assessing the efficiency of these public and private investments is one of the main jobs of research administrators; for these assessments citation bibliometrics is the most convenient tool. Evaluative bibliometrics was proposed almost 50 years ago by Francis Narin (Narin, 1976) and citation-based metrics are now used by the most reputed institutions; for example, the share of top 10% or 1% cited publications is used in different editions of documents by the OECD (https://www.oecd.org/sti/inno/scientometrics.htm), the National Science Board of the USA (e.g. National-Science-Board, 2016), the European Commission (e.g. European-Commission, 2020), and the CWTS of the University of Leiden (e.g. https://www.leidenranking.com/ranking/2022/list), among others institutions. As may be expected from their extensive use, the academic literature supporting the appropriateness of citation indicators for the evaluation of research performance is huge (reviewed by Aksnes et al., 2019; Waltman, 2016).

*1.1. The apparent puzzling case of Japan*

Despite this extensive and well-documented evidence supporting the appropriateness of citation-based metrics for research assessment, there is also overwhelming evidence showing that current bibliometric evaluations fail in the case of Japan (reviewed by Pendlebury, 2020). Its high technological and scientific level and frequent Nobel Prizes are in contradiction with its low bibliometric evaluations For example, over many years, less than 1% of the scientific publications from Japan have reached the top 1% of the world most cited publications (European Commission, 2020, Figure 6.1-8; National Science Board, 2022, Figure 23), a proportion much lower than for leading research countries and indeed at the level of developing countries. The same occurs at the university level; in the Leiden Ranking, over many years, universities such as Tokyo, Kyoto, or Osaka, have had less than 1% of their publications among the top 1% of the world's most cited publications ("All sciences" field). Remarkably, several Japanese universities with Nobel laureates, in their best year, had only 0.7% of their publications in the global top 1% of most cited publications.

These evident contradictions between citation-based metrics and reality have been investigated, and several explanations have been advanced. Regarding the comparison with the number of Nobel Prizes, it has been proposed that "even though the number of Nobel laureates in a country and citation impact is used as indicators (proxies) for measuring the quality of research, they appear to measure different aspects of quality" (Schlagberger et al., 2016, p. 731). The failure of citation-based metrics has also been explained by factors that influence citations, such as "modest levels of international collaborations," "low levels of mobility", and "Japan's substantial volume of publications in national oriented journals that have limited visibility and reduced citation opportunity" (Pendlebury, 2020, p.134).

However, these explanations are not convincing. In the case of the number of Nobel laureates, even if it were accepted that this measures a different aspect of quality than citation-based metrics, the question that arises is why Japan is different from other advanced countries. Indeed, the supposed factors that decrease the citations of Japanese publications do not occur at the very high citation levels that are relevant to scientific advances (Rodríguez-Navarro & Brito, 2022a). Furthermore, in the Clarivate "Hall of Citation laureates" (https://clarivate.com/citation-laureates/hall-of-citation-laureates/, accessed on 10/01/2022), Japan has 28 Citation laureates in natural sciences in the 2002–2021 period, more than Germany, which has 13, and France, which has 9 (Rodríguez-Navarro & Brito, 2022a).

A different explanation for the failure of citation-based indicators in Japan, which might also affect other countries, is that the weight of highly cited publications is concealed when the number of poorly cited publications is high; this occurs in Japan owing to the high proportion of technological research (Rodríguez-Navarro & Brito, 2022b).

## 2. REVIEW OF THE BASIS OF CITATION-BASED ASSESSMENTS

*2.1. Percentile indicators, rankings, and probabilities*

As mentioned in Section 1, the most prestigious institutions use percentile-based indicators when ranking the research success of countries and institutions. At least for country rankings, it may be assumed that they try to provide information about the contribution to the advancement of knowledge, which is based on publications of very low frequency. Thus, the challenge of scientometrics is to predict events of very low frequency on the basis of events of high or moderate frequency. Before addressing this challenge in the next section, it is worth reviewing the properties of the tools that could be used for this purpose.

Percentile indicators have outstanding properties, several of which arise from the fact that citations are lognormally distributed. The universality of this lognormal distribution (Radicchi et al., 2008) has been discussed extensively (reviewed by Golosovsky Golosovsky, 2021). Because this type of distribution applies to both world and country or institution publications, all publications belong to two lognormal distributions, one of which is part of the other. Consequently, all the publications of countries and institutions have two linked ranks, in the world and local rankings, and the double rank plot of these publications fit a power law (Rodríguez-Navarro & Brito, 2018a). Owing to this property, the number of publications from countries and institutions in different top percentiles also fit a power law (Brito & Rodríguez-Navarro, 2018). This property, along with many other advantages (Bornmann et al., 2013), makes percentile indicators the most convenient for research evaluations.

Among others, one remarkable characteristic of percentile indicators is that they allow for the creation of robust and objective quality rankings. In any activity, rankings by success or best performance are uncertain because different levels of success cannot be combined reliably into a single parameter. For example, the Research Excellence Framework 2014 in the UK (REF2014, 2011) rates publications according to four levels, which in common terms might be named: outstanding, excellent, good, and average. These terms can be applied to any activity, but the question in any activity is how to combine reliably these ratings to produce a ranking of players, for example, the question of how many good achievements are equivalent to an outstanding achievement. If A has four outstanding, two excellent and four good achievements, and B has two outstanding, six excellent and two good achievements, which ranks first? If this question were to arise in a literary or many other contests, it would not have a specific answer and but would depend on the opinion of the judges.

In contrast, research rankings of countries or institutions do not depend on opinions because the numbers of outstanding, excellent, good, and average publications of a country or institution are linked by an equation. If one replaces these ambiguous ratings with percentile ranks based on the number of citations—for example, ordering the papers by the number of citations, so outstanding could correspond to papers situated in the global top 1%, excellent could correspond to papers between the top 1% and 5%, and so on—the ranking is not ambiguous because the number of papers that a country or institution has in top percentiles is linked by a power law. In other words, in countries and institutions, the numbers of papers in any selected series of top percentiles are not random numbers and the question of which ranks first can be answered.

Using top percentile indicators, a mathematical constant derived from the exponent of the power law, $e_p$, characterizes the performance of institutions and countries (Rodríguez-Navarro & Brito, 2018b). The correct way of calculating the constant $e_p$ is by counting the number of papers in many top percentiles and fitting the power law, but when paper counts are high and statistically robust (Rodríguez-Navarro & Brito, 2019):

$$e_p \approx P_{top\ 10\%}/P \qquad (1)$$

where P is the total number of publications and $P_{top\ 10\%}$ is the number of publications in the top 10% of the world's papers in the same discipline and years ordered by their number of citations, with the most cited first.

The mathematical constant $e_p$ describes the rate at which the number of papers decreases as one considers increasingly narrow top percentiles. When the percentile decreases ten times (e.g. from the top 10% to the top 1%), the number of papers decreases less that ten times in the most efficient institutions or countries—five times for $e_p = 0.2$—and more than ten times in the least efficient institutions or countries—20 times for $e_p = 0.05$. Therefore, the mathematical constant $e_p$ is a special indicator of success because it is not related to any level of research achievements. Real success is better described by the size-independent probability of publishing highly cited papers, which is obviously different for each level, e.g., the top 10% or 1%. Mathematically, the formulas for these probabilities must be written on the basis of the constant $e_p$, but supposing that paper counts are statistically robust, they may also be written on the basis of the $P_{top\ 10\%}/P$ ratio (Eq.1). The formulas are the following:

Probability that a highly cited paper is in top percentile $x$ $\approx (P_{top\ 10\%}/P)^{(2-\lg x)}$ (2)

Expected number of such highly cited paper $\approx P \cdot (P_{top\ 10\%}/P)^{(2-\lg x)}$ (3)

where P can also be written also $P_{top\ 100\%}$.

In the real world, deviations of the $P_{top\ 10\%}/P$ ratio from the $e_p$ calculated by fitting the data point occur; notably, the deviations for China and Japan are different from those for the EU and the USA (Rodríguez-Navarro & Brito, 2019). However, for simplicity, henceforth we use the $P_{top\ 10\%}/P$ ratio instead of the constant $e_p$, assuming that this introduces an inaccuracy that can be tolerated for our purposes.

The conclusion that can be drawn from this analysis is that rankings of institutions or countries based on their $P_{top\ x\%}/P$ ratios do not vary with the value of $x$; they are independent of the top percentile selected (Rodríguez-Navarro & Brito, 2021). Although rankings do not vary, the differences between countries based on the $P_{top\ x\%}/P$ ratios increase as one considers increasingly narrow top percentiles. Thus, when the statistics are robust, the proportion of top 1% most cited papers is the square of the proportion of top 10% most cited papers (Eq. 2). Similarly, the probability of one exceptional publication that occurs only once out of 10,000 is equal to the $P_{top\ 10\%}/P$ ratio raised to the power of 4, and the expected number of these infrequent publications is equal to this probability multiplied by the number of publications. This forms the basis of the excellent reliability of percentile indicators for ranking countries and institutions.

The basic idea regarding percentile-based evaluations is that each top percentile corresponds to a certain level of stringency, which can eventually be determined by comparison with an independent method of evaluation. For example, in the UK Research Excellence Framework 2014 (REF2014), in the Chemistry Unit of Assessment, the 4* peer review level corresponds to the top 2.8 percentile, while in the Economic and Business Unit of Assessment, the 4* peer review level corresponds to the top 9 percentile (Rodríguez-Navarro & Brito, 2020).

*2.2. Calculating what counts by counting what can be counted*

Abramo and D'Angelo (Abramo & D'Angelo, 2016b) brilliantly raise the key challenge of scientometrics by reminding us that "*not every thing that can be counted counts and not every thing that counts can be counted*" (Cameron, 1963).

The frequency of the research achievements that produce the advance of science and result in Nobel Prizes is very low and, in principle, should be compare with citation-based metrics of a similar frequency, such as very highly cited papers (Rodríguez-Navarro, 2011) or Clarivate Citation laureates (https://clarivate.com/citation-laureates/, accessed on 10/01/2022). In contrast, a comparison of these infrequent achievements with the number of much more frequent publications might be misleading. In Kuhn's terms (Kuhn, 1970), it might be said that Nobel Prizes and very highly cited papers are indicators of revolutionary science while more common bibliometric indicators are indicators of normal science.

Revolutionary science can be predicted from normal science because top percentile indicators predict rare events by counting frequent events (Section 2.1). For example, the number of top 10% most cited publications, which is used extensively, implies 1 out of 10 publications. In contrast, publications that report Nobel Prizes achievements or are considered in Clarivate Citation laureates occur with a very low frequency, perhaps in 1 out of 10,000 or more publications. In most activities, the numbers of successful events at such different levels of stringency are independent, and the most frequent cannot be used to predict the least frequent. For example, Nobel laureates in literature cannot be predicted by counting the number of average novels; in club of novelists, the number of average novels published by the members of the club does not predict the probability that a member of the club will write an exceptional novel.

In research, publications that push the boundaries of knowledge cannot be counted with statistical reliability in most cases. Landmark publications occur with a frequency lower than 0.02% among world publications (Bornmann et al., 2018). Let us suppose a frequency of 0.02% and consider a country or institution whose research efficiency is equal to that of world research and that publishes 1000 papers per year. In this case, the expected number of annual landmark papers would be 0.2 and a robust statistical analysis of its research performance would probably require to count the number of publications during a period of 50 years. In a less efficient country or institution, this period would be even longer. These difficulties with counting are frequent, even at less stringent levels. For example, in the Leiden Ranking 2022, field of "Physical sciences and engineering," period 2017–2020, $P_{top\ 1\%}$ is counted with statistical reliability for less than half of the 1231 universities in the ranking.

If percentile indicators had not been applied in research evaluations, only normal research might be evaluated by statistically robust citation-based metrics. Using percentile indicators, the revolutionary research that pushes the boundaries of knowledge can also be evaluated by calculating the probability or frequency of these breakthrough publications (Section 2.1). However, this evaluation requires a certain consensus regarding the specific level of stringency.

An example taken from the Leiden Ranking 2022, field of "Physical sciences and engineering" illustrates these considerations. Two universities, Stanford University and the Universidade de São Paulo, published a similar number of papers in the 2017–2020 period (3708 and 3681, respectively), but as the top percentile considered narrows, the plots of the number of papers from these two universities diverge (Figure 1). Performing the assessment using all publications ($P_{top\ 100\%}$) both universities are similar but when considering $P_{top\ 0.01\%}$, Stanford University surpasses the Universidade de São Paulo by a factor of 47.

At the level of revolutionary research, the number of Nobel Prizes can be used as a control, facilitating the interpretation of Figure 1. Although the number of Nobel Prizes is an indicator that should be used with precaution because of its low frequency, it is the gold reference of scientific excellence (Charlton, 2007; Schlagberger et al., 2016). Stanford University has had more than 30 Nobel laureates in the sciences, including five in the present century in chemistry or physics, while the Universidade de São Paulo has not had any. Furthermore, in Clarivate's "Citation laureates 2022" in natural sciences (https://clarivate.com/citation-laureates/, accessed on 10/01/2022), there are two professors

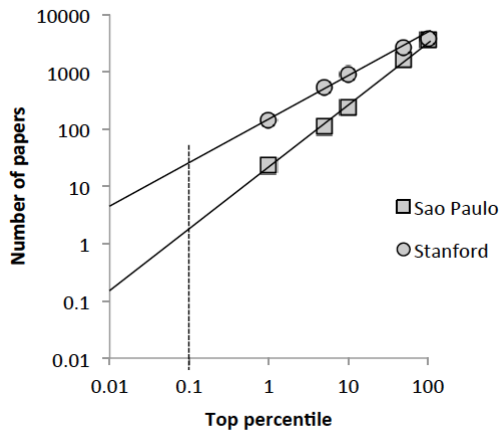

Figure 1. Number of publications in top percentiles in the universities of São Paulo and Stanford. Data taken from the Leiden Ranking, field of "Physical sciences and engineering," period 2016–2019, fractional counting

from Stanford University, while in the "Hall of Citation laureates" (https://clarivate.com/citation-laureates/hall-of-citation-laureates/, accessed on 10/01/2022) from 2002 to the present, there are no Brazilian researchers. These simple considerations suggest that, although the Universidade de São Paulo is a respectable university, its contribution to the advancement of science is far removed from the contribution of Stanford University. Certainly, this difference is not revealed by the moderate differences in $P_{top\ 10\%}$ or $P_{top\ 1\%}$; a more stringent indicator is thus needed (Figure 1).

The misleading use of an inappropriate percentile can also be deduced by focusing on productivity in economic terms as used by Abramo and D'Angelo (2016a, and other references therein). Although we could not find the actual investment in research by the Universidade de São Paulo, it seems clear that it is much smaller (https://fapesp.br/en/about, accessed on September 28, 2022) than for Stanford University (https://facts.stanford.edu/research/, accessed on September 28, 2022). Therefore, if productivity were to be calculated on the basis of $P_{top\ 10\%}$ or $P_{top\ 1\%}$, Stanford University would obtain a misleadingly poor evaluation.

It has been suggested that Nobel laureates and citation impacts might measure different aspects of research quality (Schlagberger et al., 2016), but according to the comparison presented above, another interpretation is more likely. Nobel laureates, as well as Clarivate Citation laureates, are characterized by very highly cited publications in the heavy tail of the citation distribution, far from the region where most citation-based indicators are calculated. This implies that the number of Nobel laureates and citation impact indicators do not measure different aspects of quality; rather, they measure quality at two different levels of stringency.

In summary, the conclusion that can be drawn from the considerations above is that an assessment of the contribution of countries and institutions to the advancement of knowledge cannot be performed directly by counting the number of publications that are very infrequent. Rather, such an assessment is possible by calculating the probability or expected frequency of publishing at more stringent percentiles than those in which counting is statistically reliable. There is not a single top percentile for this assessment, but previous studies have shown that a convenient top percentile is 0.01. In other words, a like-for-like indicator for the contributions to the advancement of knowledge would be $P_{top\,0.01\%}$ (Brito & Rodríguez-Navarro, 2018).

Returning to the problem of the evaluation of Japan, it is worth noting that the use of $P_{top\,0.01\%}$ for the research evaluation does not improve its misleading evaluations; the root of this problem lies in the low $P_{top\,10\%}/P$ ratio. Apparently, for most countries, $P_{top\,0.01\%}$ can be calculated from the $P_{top\,10\%}/P$ ratio, but not in Japan.

*2.3. Research is not only for pushing the boundaries of knowledge*

A high proportion of the research performed in many countries and institutions is focused on pushing the boundaries of knowledge. However, research may serve for other purposes as well. For example, in some higher education institutions, research is centered on supporting teaching roles while, at some institutions, research may be focused on incremental innovations. In these two cases, the expected number of citations to their publications cannot be very high—it is unlikely that the results of a master's thesis or of an incremental innovation will be highly cited. At institutions in which research is focused only on master's theses or incremental innovations, research assessments may be performed at low citation levels or wide top percentiles (e.g. $P_{top\,50\%}$), which can be achieved by counting. In contrast, $P_{top\,0.01\%}$ needs to be calculated, and the problem with Japan arises in this calculation because a large proportion of its research is focused on incremental innovations (Rodríguez-Navarro & Brito, 2022b).

## 3. RATIONALE AND AIM OF THIS STUDY

*3.1. Inaccuracies in the calculation of $P_{top\ 0.01\%}$*

The exceptional characteristic of percentile indicators is that they allow for the assessment of revolutionary research by counting citations at the level of normal research, but this procedure fails in Japan. The issue that then arises is if this occurs only in Japan or in many other countries when there are several groups with their research focused on different objectives.

A numerical example illustrates this problem. Let us suppose that an institution or country has two groups of researchers who publish 500 papers each in exactly the same research field, with the research of one group focused on the progress of knowledge and the research of the other addressing incremental innovations. For the first group, 20% of their papers are in the top 10% of the world's most cited papers, while in the other group, none of the papers are in this top 10%. The probabilities of each of these two groups of researchers publishing a paper in the top 0.01% most cited papers is calculated from the $P_{top\ 10\%}/P$ ratio (Eq. 2), with the values being 0.0016 (= $0.2^4$) for the first group and zero for the second group. Because each group publishes 500 papers, the expected number of top 0.01% most cited papers is 0.8 and zero (Eq. 3). Consequently, the sum of both numbers is 0.8, which is also the expected number for the whole institution. In contrast, if we treated the two groups as a whole, we would aggregate the papers of the two groups before the calculation. Then the total number of papers would be 1000, the $P_{top\ 10\%}/P$ ratio would be 0.1, and the probability for a paper to be in the top 0.01% most cited would be 0.0001. In consequence, the calculated number of top 0.01% most cited papers would be 0.10, eight times lower than the real number.

Although this example is fictitious and probably not realistic, it alerts us to a possible inaccuracy when research assessments for countries and institutions are focused on revealing the contribution to the advancement of science but on the basis of indicators calculated from citation counts for the whole country or institution.

*3.2. Central hypothesis*

If one defines units of research as groups of researchers with the same research aim and having a similar probability of making discoveries, nearly all countries or institutions will be made up of several or many different units of research. Consequently, for most countries and institutions, the expected number of papers in the top 0.01% most cited papers, can be calculated by two procedures: (i) adding the results of Eq. 3 obtained from $P_{top\ 10\%}$ and P for each units of research (CFAL method, calculating first and adding later) or (ii) aggregating all the publications of the units of research, which is the same as counting $P_{top\ 10\%}$ for the whole institution or country, and using the resulting $P_{top\ 10\%}$ and P in Eq. 3 (AFCL method, aggregating first and calculating later). Of these two procedures, only the first is correct, while the second produces a misleadingly low result.

Considering *n* elemental units of research in a country or institution, this conclusion can be expressed by the following inequation:

$$(\sum_{k=1}^{n} Pk) \cdot \left(\frac{\sum_{k=1}^{n} Ptop\ 10\%k}{\sum_{k=1}^{n} Pk}\right)^{(2-\lg x)} < \sum_{k=1}^{n} \left(Pk \left(\frac{Ptop\ 10\%k}{Pk}\right)^{(2-\lg x)}\right) \qquad (4)$$

where *Pk* and $P_{top\ 10\%}k$ correspond to the values of these indicators in the research unit *k*. On the left-hand side, the expected number of highly cited papers is calculated by using the AFCL method, whereas on the right, the expected number of highly cited papers is calculated using the CFAL method. This inequation holds when (2 - lg *x*) > 1.

Ineq. 4 implies that, given the existence of different units of research in most universities, all countries, and all aggregations of countries, e.g. the EU, the $P_{top\ 10\%}$/P ratio is not an accurate indicator of the contribution to the advancement of science, because Eq. 2 and 3 do not hold.

*3.3. A simulation approach*

Ineq. 4 can be tested by using the data published for universities in the Leiden Ranking, taking universities as the units of research defined in the previous section. This test, however, is unlikely to reveal the actual size of this problem, because in most cases, universities will be made up of groups with different probabilities of making important discoveries.

Therefore, to address the hypothesis in a systematic manner, we used a simulation approach with synthetic series of numbers, simulating series of papers whose citations are lognormally distributed (Section 2.2). Although in practice there are cases with deviations from the lognormal distribution (Waltman et al., 2012), this does not affect to our analyses, which can be centered in the most common cases in which citations are lognormally distributed.

Because the causes that give rise to lognormal distributions (e.g. Limpert et al., 2001; Redner, 2005) are the same at all levels of aggregation, it can be concluded that, regarding the citation counts of research papers, aggregation of lower-level lognormal distributions will produce another lognormal distribution. This most likely occurs because the lower-level citation distributions that make up higher-level citation distributions—for example, institutions that make up countries—are not random lognormal distributions but rather distribute responding to a certain order. This is similar to what has been described for the number of papers in top percentiles: they are not random numbers (Section 2.2). Considering this restriction, our simulations are not random: they respect certain orders.

*3.3. Summary, hypothesis, and aims*

The contribution of countries and institutions to the advancement of science is basic information in research policy, but such advances are reported in publications that are very highly cited and very infrequent. Because of this infrequency, the number of such publications, for example, $P_{\text{top } 0.01\%}$ after normalization, cannot be counted in most cases. Current theoretical issues support that country and institution rankings based on $P_{\text{top } 10\%}$ or $P_{\text{top } 1\%}$ mimic the rankings based on $P_{\text{top } 0.01\%}$. Therefore, $P_{\text{top } 10\%}$-based country rankings, which can be rigorously determined, are almost identical to the $P_{\text{top } 0.01\%}$-based country rankings (Rodríguez-Navarro & Brito, 2021), which reveal contributions to the advancement of science. However, in contrast to this general property, for Japan, the $P_{top\ 10\%}$-based evaluation does not mimic the evaluation based on very highly cited publications (Section 1.1).

The hypothesis of this study is that a restriction should be added to the general property of the identity of percentile rankings, that it is true only when all the researchers in countries and institutions have similar research aims and efficiency. Japan would not satisfy this criterion, and it seems possible that the same will occur for other countries and institutions. This implies that $P_{top\ 10\%}$ or $P_{top\ 1\%}$ are inaccurate indicators for the evaluation of the contribution to the advancement of science, despite being comparatively correct in most cases.

Under these circumstances, our study aims to investigate the inaccuracy when using the aggregated values of P and $P_{top\ 10\%}$ to calculate the contribution of countries and institutions to the advancement of science in a specific topic. Assuming that a country or institution is made up of groups of researchers with different probabilities of publishing very highly cited papers, we calculated the expected $P_{top\ 0.01\%}$: (i) using P and $P_{top\ 10\%}$ for the country or institution (the AFCL method) and (ii) adding the values calculated from the P and $P_{top\ 10\%}$ of the independent research groups (CFAL method).

## 4. MATERIAL AND METHODS

In the first part of this study, we tested the hypothesis presented in Ineq. 4, considering universities as the units of research that make up a country. For this purpose, we calculated $P_{top\ 0.01\%}$ of countries using the P and $P_{top\ 10\%}$ values of universities, applying the CFAL and AFCL methods. The university data were obtained from the Leiden Ranking 2022, for the period 2016–2019, for the field of "Physical sciences and engineering," and fractional counting.

In the second part of this study, we used a simulation approach. We suppose that the world research in a certain field is produced by 400 units of research, each of them publishing 200 articles per year. We thus generated 400 primary series, each with 200 lognormal distributed numbers, which mimic the number of citations of each paper. Next, we combined these 400 primary series, which simulated 80,000 world publications, and ordered them starting with the highest number. In this synthetic world series, we recorded the number ranked in position 8000 (top 10%). To simulate institutions or countries, we selected 20 series from the 400 primary series. The selection of these 20 series was performed according to different approaches that are described in the Results section.

The procedure to generate these series of random numbers that are lognormal distributed has been described previously (Rodríguez-Navarro & Brito, 2018a; Thelwall, 2016c). To select the values of $\mu$ and $\sigma$ for these series, we studied the lognormal distributions of citations in selected topics, such as lithium batteries, solar or photovoltaic cells, graphene, electrochemistry, energy and fuels, cancer, stem cells, immunity, CRISPR, and a few others in top and bottom universities in the Leiden Ranking. The value of $\mu$ was obviously dependent on the citation window, because this window changes the number of citations, but in our tests, the ratio between the maximum and minimum values of $\mu$—for top and bottom universities—in the same topic and citation window was always around 2:1. In the case of $\sigma$, its value was always around 1.0, in accordance with previous studies (Radicchi et al., 2008; Rodríguez-Navarro & Brito, 2018a; Thelwall, 2016a; Viiu, 2018), but varying depending on the institution and topic; for the same institution and topic, $\sigma$ varied very little and was almost independent of the variation of $\mu$, which depended on the counting window.

On the basis of this information, we generated the 400 synthetic primary series with $\mu$ values from 4.0 to 2.0, a constant $\sigma$ value of 1.1, and 200 numbers. To program the variation of $\mu$, we used a lineal approach, decreasing $\mu$ linearly from 4.0 to 2.0 (the use of a power law decrease did not change the results). Each series is characterized by a $P_{top\ 10\%}/P$ ratio with reference to the simulated world series; the distribution of this ratio was not lineal but approximately fits a quadratic function (Figure 2).

The generated synthetic series are continuous, and for most bibliometric purposes these series are discretized because the numbers of citations are integers (Rodríguez-Navarro & Brito, 2018a; Thelwall, 2016c). However, "it is reasonable to use the continuous approximation in order to be able to mathematically analyze citation indicators" (Thelwall, 2016b, p. 872), and in this study, we did not discretize the series. After discretization, there are many simulated publications with the same number of citations in the world and in the

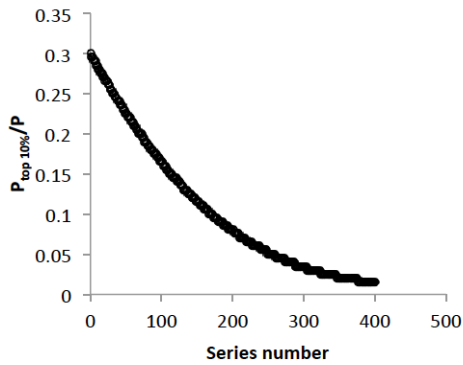

Figure 2. Distribution of the $P_{top\,10\%}/P$ ratio of the 400 primary series used in this study with reference to the combined series

20 combined series; these tied publications make it difficult to determine the number of top 10% publications (Waltman & Schreiber, 2013). Without discretization, the $P_{top\,10\%}$ values for the simulated institutions or countries can be established unequivocally, because there are no simulated publications with the same number of citations.

The calculations were based on the following two approaches: For the CFAL method, the $P_{top\,10\%}$ of each primary series was determined and divided by 200 ($P_{top\,10\%}/P$), then this quotient was raised to the power of four and multiplied by 200 (Eq. 3). The resulting $P_{top\,0.01\%}$ of the 20 series were added to obtain the $P_{top\,0.01\%}$ of the combined 20 series. For the AFCL method, the 20 series were combined, and the resulting series was ordered starting with the highest number. In these combined series, $P_{top\,10\%}$ was determined and divided by 4000 ($P_{top\,10\%}/P$); the resulting quotient was raised to the power of four and multiplied by 4000 to obtain the $P_{top\,0.01\%}$.

The series simulating world publications with 80,000 numbers and all series obtained from the aggregation of 20 primary series were lognormally distributed (Kolmogorov-Smirnov tests, $p > 0.15$).

## 5. RESULTS

*5.1. Countries as aggregation of universities*

As already explained in Section 3.2, we used the Leiden Ranking for an empirical approach to test Ineq. 1, considering countries as the aggregation of their universities. We then calculated the expected $P_{top\ 0.01\%}$ with the CFAL and AFCL methods (Section 3.3).

Table 1. Truncated table of the expected $P_{top\ 0.01\%}$ for US universities calculated by the CFAL method and the cumulative values. The last row shows the calculation by the AFCL method[a]

| Rank | University | P | $P_{top\ 10\%}$ | $P_{top\ 10\%}/P$ | $P_{top\ 0.01\%}$ | Cumulative $P_{top\ 0.01\%}$ |
|---|---|---|---|---|---|---|
| 1 | MIT | 4803 | 1149 | 0.24 | 15.73 | 15.7 |
| 2 | Harvard University | 3089 | 802 | 0.26 | 14.05 | 29.5 |
| 3 | Stanford University | 3729 | 920 | 0.25 | 13.84 | 43.3 |
| 4 | UC, Berkeley | 3475 | 788 | 0.23 | 9.20 | 52.2 |
| 5 | CALTECH | 2889 | 613 | 0.21 | 5.84 | 57.3 |
| 6 | Northwestern University | 2675 | 559 | 0.21 | 5.10 | 61.8 |
| 7 | Princeton University | 2490 | 521 | 0.21 | 4.76 | 66.1 |
| 8 | University of Chicago | 1463 | 337 | 0.23 | 4.14 | 70.3 |
| 9 | Yale University | 1345 | 309 | 0.23 | 3.77 | 74.0 |
| 10 | UC, Los Angeles | 2486 | 486 | 0.20 | 3.64 | 76.9 |
| 11 | UC, Santa Barbara | 1917 | 389 | 0.20 | 3.26 | 79.8 |
| 12 | Columbia University | 1583 | 335 | 0.21 | 3.17 | 82.9 |
| 13 | Cornell University | 2002 | 391 | 0.20 | 2.91 | 89.4 |
| 14 | UC, San Diego | 2270 | 414 | 0.18 | 2.51 | 89.6 |
| 15 | UT, Austin | 3315 | 548 | 0.17 | 2.47 | 89.7 |
|  |  |  |  |  |  |  |
| 191 | Portland State University | 169 | 11 | 0.06 | 0.0026 | 145.9 |
| 192 | UNC, Greensboro | 63 | 5 | 0.08 | 0.0026 | 145.9 |
| 193 | UT, Health S Cent, San Antonio | 24 | 2 | 0.10 | 0.0022 | 145.9 |
| 194 | Thomas Jefferson University | 25 | 2 | 0.10 | 0.0022 | 145.9 |
| 195 | Rush University | 18 | 2 | 0.09 | 0.0011 | 145.9 |
| 196 | East Carolina University | 56 | 4 | 0.06 | 0.0009 | 145.9 |
| 197 | Florida Atlantic University | 174 | 8 | 0.05 | 0.0008 | 145.9 |
| 198 | Univer of Alaska, Fairbanks | 120 | 6 | 0.05 | 0.0008 | 145.9 |
| 199 | Unifor Ser U of Health Sciences | 10 | 1 | 0.07 | 0.0003 | 145.9 |
| 200 | Loyola University Chicago | 38 | 2 | 0.04 | 0.0001 | 145.9 |
|  |  |  |  |  |  |  |
|  | AFCL method | 171722 | 25881 | 0.15 | 88.61 |  |

[a] The table shows the 15 first and 10 last universities of the complete table of the country. The values of P and $P_{top\ 10\%}$ were taken from the Leiden Ranking, field of "Physical sciences and engineering," period 2016-2019, fractional counting

We first studied the USA, where there are 200 universities whose $P_{top\ 10\%}/P$ ratios vary from 0.24 to 0.04. The expected values of $P_{top\ 0.01\%}$ calculated by the CFAL and AFCL methods were 150 and 89, respectively (Table S1; Table 1 shows a sample of 25 universities). Ordering the universities in descending order of their $P_{top\ 0.01\%}$, the cumulative CFAL values show that the first 13 universities are sufficient to account for the value of the expected $P_{top\ 0.01\%}$ when calculated by the AFCL method (Table S1; Figure 3). A similar

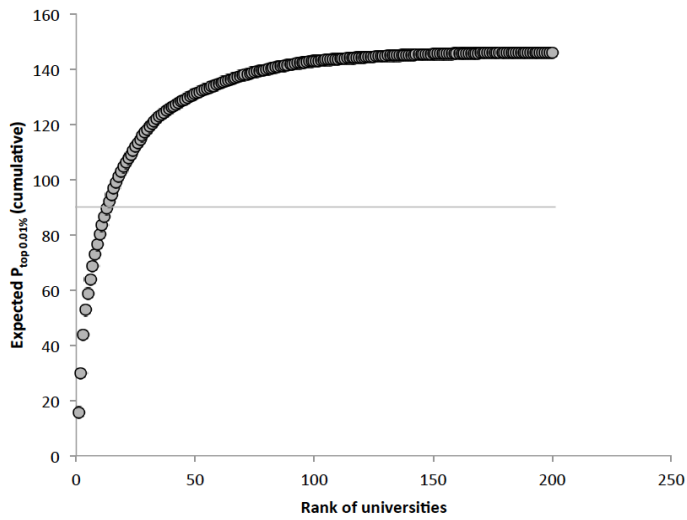

Figure 3. Cumulative plot of the expected $P_{top\ 0.01\%}$ in US universities using the CFAL method. Calculations from P and $P_{top\ 10\%}$ in the Leiden Ranking, field of Physical sciences and engineering, period 2016-2019, and fractional counting. The horizontal line shows the $P_{top\ 0.01\%}$ calculated by the AFCL method

study for the UK and Germany (Figure 4) showed that the CFAL and AFCL calculation methods produce more similar results than for the USA, with a difference of approximately 15% for the UK and Germany versus 67% for the USA (Table 2).

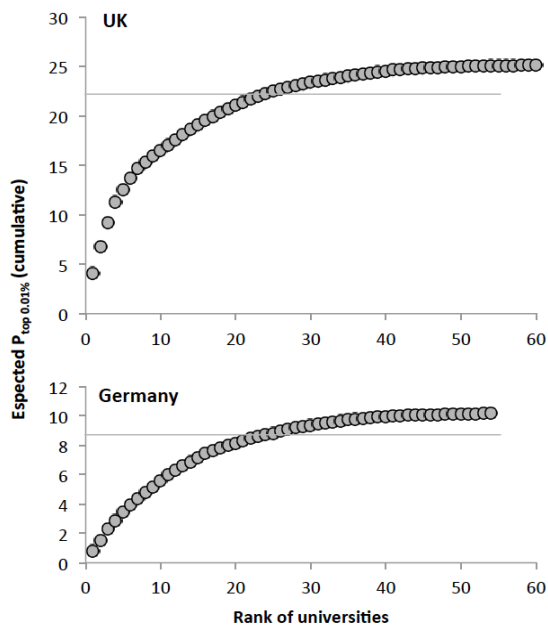

Figure 4. Cumulative plot of the expected $P_{top\ 0.01\%}$ in UK and Germany universities using the CFAL method. Calculations from P and $P_{top\ 10\%}$ in the Leiden Ranking, field of Physical sciences and engineering, period 2016-2019, and fractional counting. The horizontal line shows the $P_{top\ 0.01\%}$ calculated by the AFCL method

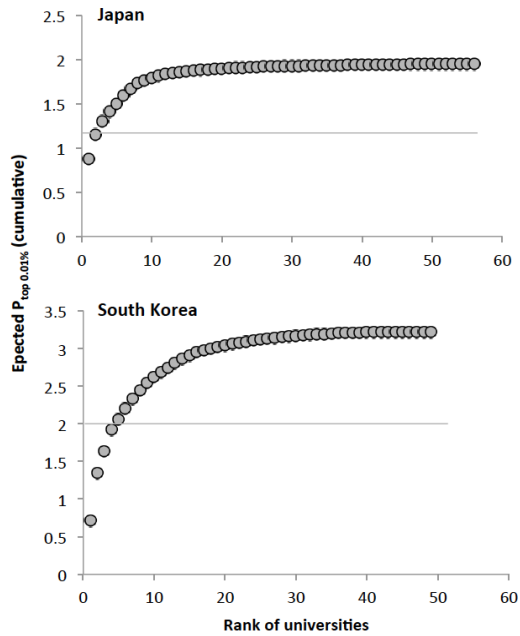

Figure 5. Cumulative plot of the expected $P_{top\,0.01\%}$ in Japan and South Korea universities using the CFAL method). Calculations from P and $P_{top\,10\%}$ in the Leiden Ranking, field of Physical sciences and engineering, period 2016-2019, and fractional counting. The horizontal lines show the $P_{top\,0.01\%}$ calculated by the AFCL method

Next, we studied Japan and South Korea; Figure 5 shows the cumulative plots of $P_{top\,0.01\%}$ for these countries. Similarly to the USA, the $P_{top\,0.01\%}$ values calculated by the CFAL method were 66% and 60% higher than those calculated by the AFCL method (Table 2). In japan, when using the CFAL method, only two universities accounted for 98% of the total value of the expected $P_{top\,0.01\%}$ when calculated by the AFCL method.

Table 2. Expected $P_{top\,0.01}$ country values calculated by the AFCL and CFAL methods considering universities as units of research[a]

| Country | AFCL method | | | | CFAL method | |
|---|---|---|---|---|---|---|
| | P | $P_{top\,10\%}$ | $P_{top\,10\%}/P$ | $P_{top\,0.01}$ | P | $P_{top\,0.01}$ |
| USA | 171722 | 25881 | 0.151 | 88.6 | 171722 | 146 |
| UK | 53527 | 7632 | 0.143 | 22.1 | 53527 | 25.1 |
| Germany | 54235 | 6076 | 0.112 | 8.54 | 54235 | 10.1 |
| Japan | 51927 | 3584 | 0.069 | 1.18 | 51927 | 1.94 |
| South Korea | 52067 | 4089 | 0.079 | 1.98 | 52067 | 3.21 |
| China | 432182 | 46648 | 0.108 | 58.6 | 432182 | 77.7 |
| India | 46786 | 3456 | 0.074 | 1.39 | 46786 | 1.83 |

[a] The values of P and $P_{top\,10\%}$ were taken from the Leiden Ranking, field of "Physical sciences and engineering," period 2016-2019, and fractional counting

Finally, we studied China and India; in both of these countries, the results of the CFAL method were 32% higher than those obtained by the AFCL method (Table 2). This comparison of Japan and India is illustrative as the expected $P_{top\,0.01\%}$ calculated by the AFCL method suggests that India contributes more than Japan to the advancement of knowledge (1.39 versus 1.18). The CFAL method corrects this situation a little (now, 1.94 versus 1.83), but neither of the two methods is realistic because Japan is substantially ahead of India in terms of technological level, Nobel Prizes (Schlagberger et al., 2016), number of very highly cited publications (Rodríguez-Navarro & Brito, 2022a), or Clarivate "Citation laureates."

*5.2. Simulation approach*

The conclusion that can be drawn from these results is that a reliable indicator that describes the contribution to the advancement of knowledge in a certain field must consider the units of research and use the CFAL method for the calculation of the expected number of publications in the top percentile that is selected. As far as we know, this approach has never been applied and there is not a clear procedure for the identification of these units of research. Therefore, to investigate the differences that might emerge when calculating $P_{top\,0.01\%}$ by the CFAL versus AFCL methods, we used a simulation approach. The advantage of this method is its flexibility to investigate very different models of combinations of units of research in institutions and countries.

As described in Section 4, we used 400 synthetic series of 200 lognormal distributed numbers simulating 400 units of research, in which the efficiency for the contribution to the advancement of knowledge varied from the level of the most important universities—Harvard, Stanford, MIT, etc.—to the level of the bottom universities in the Leiden Ranking. To simulate institutions or even countries we combined 20 of these primary series, each simulating a single unit of research. To obtain information about how to combine these series, we studied the distribution of the $P_{top\,10\%}/P$ ratios of US universities. The USA is a large country with a common language and high mobility of researchers; it is possible that universities behave as units of research with low variability of the $P_{top\,10\%}/P$ ratios among research groups in the same university. Figure 6 shows that the distribution is of sigmoid type, with most of the universities having $P_{top\,10\%}/P$ ratios between 0.2 and 0.08. For comparison, we obtained the distribution of the same ratio for UK universities. The distributions of the $P_{top\ 10\%}/P$ ratios in US and UK universities were similar, although the sigmoid distribution for the UK universities was smoother (Figure 6).

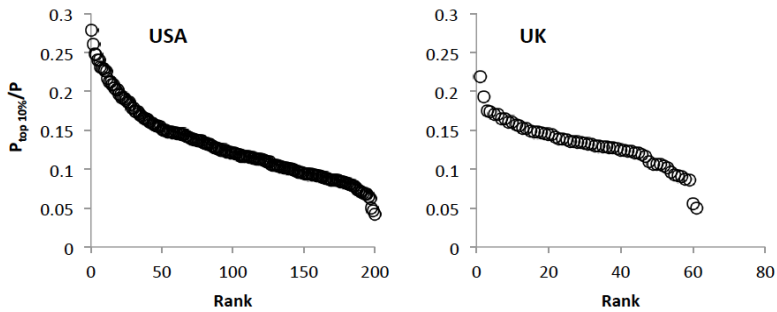

Figure 6. Distribution of the $P_{top\ 10\%}/P$ ratio versus rank in USA and UK universities. Data taken from the Leiden Ranking, field of Physical sciences and engineering, period 2019-2019, and fractional counting

Next, we combined 20 primary series in which the $P_{top\ 10\%}/P$ ratios follow four different distribution types (Figure 7): (i) linear, (ii) similar to the USA distribution, (iii) similar to the UK distribution, and (iv) split type, in which most simulated groups have either high or low $P_{top\ 10\%}/P$ ratios. This last combination simulated institutions in which approximately one-third of the units of research pursue the advance of knowledge, another third of the units of research pursue technical innovations, while the last third simulates the transition. This model resembles what might occur in Japan (Rodríguez-Navarro & Brito, 2022b).

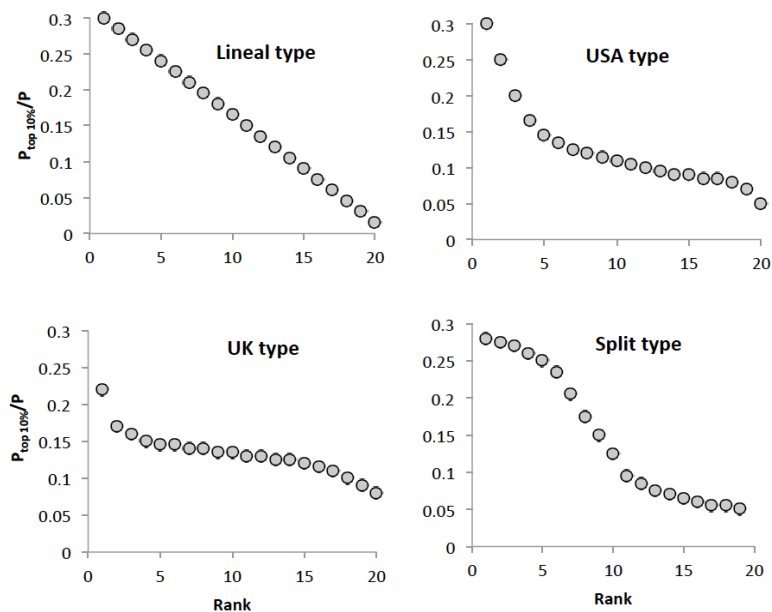

Figure 7. Distribution of the $P_{top\ 10\%}/P$ ratio versus rank in the four types of 20-series aggregations used for testing the CFAL and AFCL methods

Table 3 summarizes the values of $P_{top\ 0.01\%}$ as calculated by the CFAL and AFCL methods. As stated by Ineq. 4, the CFAL values are always higher than the AFCL values. The UK-type distribution of $P_{top\ 10\%}/P$ ratios shows the lowest difference of 60%, while the differences in the other three cases vary from two for the lineal type, to four times higher for the US and split types.

Table 3. Simulation of countries or institutions with synthetic series. $P_{top\ 0.01\%}$ calculated by the AFCL and CFAL methods for the series that combine 20 primary series. Models of combinations are shown in Figure 6

| Series | AFCL | | CFAL |
|---|---|---|---|
| | $P_{top\ 10\%}/P$ | $P_{top\ 0.01\%}$ | $P_{top\ 0.01\%}$ |
| Lineal | 0.15 | 2.15 | 7.32 |
| USA type | 0.12 | 0.85 | 3.30 |
| UK type | 0.13 | 1.08 | 1.69 |
| Split type | 0.14 | 1.57 | 6.48 |

*5.3. Comparison of Japanese and US universities*

These results demonstrate that the use of the AFCL method may result in a large underestimation of the contribution to scientific advances. Currently, there is no procedure to apply the CFAL method to Japanese universities for comparison with the common AFCL method, which appears to be highly misleading. Therefore, to provide some information regarding this underestimation, we compared Japanese and US universities of similar scientific level.

In terms of Nobel Prizes in Chemistry, Physics, and Physiology/Medicine, the Japanese universities of Tokyo, Kyoto, Nagoya, Tsukuba, and Hokkaido have a level similar to that of the US universities of Cornell, Yale, Colorado Boulder, Rice and Columbia (Schlagberger et al., 2016). Therefore, it might be expected that these Japanese and US universities will contribute similarly to the advancement of science, and that their $P_{top\ 0.01\%}$ calculated using the CFAL method should be similar. Assuming that the US universities are almost perfect units of research (Section 5.2), we may take these universities as a reference for the result of the CFAL method. This comparison of Japanese and US universities seems reasonable because the selected Japanese universities have more Clarivate Citation laureates than the US universities (results not shown), which rules out any biliometric advantage of US universities.

Using the data in the Leiden Ranking (Physical sciences and engineering, period 2011–2014), $P_{top\ 0.01\%}$ for US universities varies from 2.2 to 5.1 with a mean of 3.1; in contrast, for Japanese universities, the variability was very large from 0.05 to 0.91 with a mean of 0.33 (Table 4). The $P_{top\ 10\%}/P$ ratios also show an approximately tenfold advantage for US universities.

Although the high variability only allows for a rough numerical comparison to be made, the general conclusion is that, for similar number of Nobel Prizes, the $P_{top\ 0.01\%}$ indicator calculated for US universities may be tenfold higher than for Japanese universities. This large difference provides a rough estimate of the difference between the results of calculating $P_{top\ 0.01\%}$ by the CFAL and AFCL methods.

Table 4. Expected $P_{top\ 0.01\%}$ and $P_{top\ 0.01\%}/P$ ratio for a selection of Japanese and US universities with similar number of Nobel laureates[a]

| University | P | $P_{top\ 10\%}/P$ | $P_{top\ 0.01\%}$ | $P_{top\ 0.01\%}/P$ |
|---|---|---|---|---|
| Nagoya | 2544 | 0.088 | 0.153 | 6.00E-05 |
| Kyoto | 5440 | 0.096 | 0.462 | 8.49E-05 |
| Tsukuba | 1338 | 0.076 | 0.045 | 3.34E-05 |
| Hokkaido | 2363 | 0.074 | 0.071 | 3.00E-05 |
| Tokyo | 6245 | 0.110 | 0.914 | 1.46E-04 |
| Cornell | 2081 | 0.188 | 2.600 | 1.25E-03 |
| Yale | 1332 | 0.222 | 3.235 | 2.43E-03 |
| Colorado Boulder | 1816 | 0.188 | 2.269 | 1.25E-03 |
| Rice | 1224 | 0.254 | 5.095 | 4.16E-03 |
| Columbia NY | 1462 | 0.196 | 2.158 | 1.48E-03 |

[a] The values of P and $P_{top\ 10\%}$ were taken from the Leiden Ranking, field of "Physical sciences and engineering," period 2016-2019, fractional counting

# 6. DISCUSSION

*6.1. The failure of some bliometric predictions in Japan can be explained*

This study aims to provide an explanation for the discrepancy between the low bibliometric evaluations of Japan and its high scientific level (Pendlebury, 2020). For example, for Japan, the gold-standard bibliometric indicator $P_{top\ 10\%}/P$ ratio, as calculated by the National Science Board (2016, Appendix Table 5-59) and the European Commission (2018, 2020), has been stable at around 0.07–0.08 throughout this century, significantly lower than the global reference of 0.1 and at the level of developing countries whose contributions to scientific advances is obviously low.

These results for Japan are obviously correct because they are counts that cannot be misleading. Furthermore, a high number of publications reporting incremental innovations with a low number of citations, which implies low $P_{top\ 10\%}/P$ ratios, is not surprising in countries with a high technological level (Rodríguez-Navarro & Brito, 2022b). The contradiction arises when the $P_{top\ 10\%}/P$ ratio is taken as an indicator of the contribution to the advancement of science. If "national science indicators for Japan present us with a puzzlement" (Pendlebury, 2020, p. 134), it is because it is assumed that national science indicators such as the $P_{top\ 10\%}/P$ ratio reflect the capacity to contribute to the advancement of science. This is a frequent assumption albeit not commonly declared; most citation-based research assessments using $P_{top\ 10\%}$ or $P_{top\ 1\%}$ are supposed to be valid for judging the contribution to the advancement of knowledge. On the basis of this assumption, the results of these assessments are sometimes compared with the number of Nobel laureates.

In the case of Japan, the $P_{top\ 10\%}/P$ ratio is apparently inconsistent with its Nobel Prize level, but there is no contradiction. The like-for-like bibliometric indicator for assessing the contribution to the advancement of knowledge is $P_{top\ 0.01\%}$, and as a general rule, this indicator cannot be obtained by counting the number of publications; it must be calculated from P and the $P_{top\ 10\%}/P$ ratio (Section 2), and the flaw lies in this calculation. Considering only universities, our results show that the $P_{top\ 10\%}/P$ ratio for Japan is lower than for Germany but similar to that for India (Table 2), whereas Japanese universities have received more Nobel Prizes than Germany, while India has not received any at all (Schlagberger et al., 2016). Most probably, if $P_{top\ 0.01\%}$ were obtained by counting, the apparent contradiction would not exist. In fact, the numbers of very highly cited publications and Clarivate Citation laureates (Section 1.1) in Japan are consistent with the number of Nobel laureates. This is a key observation because it demonstrates that the failure cannot be explained by factors that decrease citations, such as "modest levels of international collaborations," "low levels of mobility," and "Japan's substantial volume of publications in national oriented journals that have limited visibility and reduced citation opportunity" (Pendlebury, 2020, p.134).

If citations to publications are not biased in Japan with respect to other countries and $P_{top\,0.01\%}$ is as good indicator of scientific advances, the logical conclusion is that the link between the country $P_{top\,10\%}/P$ ratio and $P_{top\,0.01\%}$ fails in Japan. In other words, in Japan, a low $P_{top\,10\%}/P$ country ratio is compatible with a high scientific level, a real high $P_{top\,0.01\%}$, and frequent Nobel Prizes.

To explain this, our key finding is that, although citations to publications from a country or institution follow a single lognormal distribution, the use of the P and $P_{top\,10\%}$ values of countries and institutions may lead to erroneous conclusions regarding the probability of publishing a very highly cited paper (e.g. a Nobel class paper). The flaw occurs when the lognormal distribution arises from the aggregation of several lognormal distributions that correspond to groups of researchers that, in the same research field, do not pursue the same objectives or have the same research capacity. We used the term "units of research" to describe the aggregations of researchers that are homogeneous according to their research objectives or capacity. Each unit of research has a different probability of publishing highly cited papers, and citations to their publications follow different lognormal distributions than the other units of research. For each unit of research, the probability or expected frequency of top 0.01% papers can be correctly calculated from P and $P_{top\,10\%}$, and the method for calculating of the $P_{top\,0.01\%}$ of a country or institution is obviously the addition of the $P_{top\,0.01\%}$ values of its units of research. However, if $P_{top\,0.01\%}$ is calculated from the P and $P_{top\,10\%}$ of the whole lognormal distribution generated by the aggregation of the primary distributions, the results of such a calculation is lower than the real value (Ineq. 4).

According to these observations, a different distribution of the units of research in Japan would be sufficient to make Japan unique from other countries, as already suggested (Rodríguez-Navarro & Brito, 2022b).

In Section 3.1, we provide a simple example of a fictitious institution with only two groups of research that are very different in terms of their competitiveness or research aims and two ways to calculate $P_{top\ 0.01\%}$, i.e., the CFAL and AFCL methods. This example is incorrect because the citation distributions may not be lognormal, yet it illustrates the different results that may be obtained when using the CFAL and AFCL methods. We have also shown that, simply by using the CFAL method with Japanese universities, the bibliometric evaluation of Japan increases by 60% (Table 2). It is noteworthy that this calculation does not aim to correct the evaluation of Japan because it supposes that universities as a whole are units of research, which is not correct. This supposition may not be true in any country, and it is especially erroneous for Japan because universities such as Tokyo, Kyoto, Nagoya, or Hokkaido have Nobel Prize level (Schlagberger et al., 2016) and very low $P_{top\ 10\%}/P$ ratios. To further confirm that Japanese universities cannot be consider as units of research, we compared the $P_{top\ 10\%}/P$ ratios and calculated $P_{top\ 0.01\%}$ for some US and Japanese universities that are similarly successful in terms of Nobel Prizes, finding that the values of these parameters for Japanese universities are much lower than for US universities (Table 4). This large difference between universities that probably make similar contributions to the advancement of science strongly suggests that the units of research in US universities are less diverse than in Japanese universities.

In summary, the contribution to the progress of knowledge cannot be estimated from indicators such as $P_{top\ 10\%}$ or $P_{top\ 1\%}$ when they are counted for whole countries or institutions. In this case, their use as indicators of the contribution to the advancement of science is mathematically incorrect.

*6.2. Size of the differences between the CFAL and AFCL methods*

The real size of the deviations from reality of the AFCL method as revealed by the CFAL method, when assessing the contribution to the progress of knowledge is unknown; the deviation is also difficult to test because there is no a clear way of identifying the units of research. Given these difficulties we estimated the deviations with synthetic series.

In order for the numerical differences between the CFAL and AFCL methods to be as realistic as possible, we used the distribution of the $P_{top\ 10\%}/P$ ratios in US universities as a reference for the distribution of the $P_{top\ 10\%}/P$ ratios in the 20 series that simulate institutions or countries. As already pointed out in Section 5.2, the USA is a large country with a common language and high mobility of researchers, in which university researchers may have similar efficiencies and universities might simulate units of research. With the USA-type distribution of the $P_{top\ 10\%}/P$ ratios, the difference between the CFAL and AFCL methods was higher than 3:1 (Table 3). When researchers are distributed into two groups, viz. the split type, as proposed for Japan, with one group of researchers pursuing the advancement of knowledge and the other pursuing incremental technical innovations (Rodríguez-Navarro & Brito, 2022b), the difference was 4:1. These simulations seem realistic because the $P_{top\ 10\%}/P$ ratios of the simulations—this ratio exists only in the AFCL method—varied between 0.12 and 0.15, which are normal values in the Leiden Ranking (equivalent to $PP_{top}$ indicators of 12% and 15%).

In the comparison between US and Japanese universities (Table 4), the difference between their $P_{top\ 0.01\%}$ values was greater that in our simulations (Table 3), but the purpose of the simulations was only to show in controlled tests that the differences may be high. Certainly, in the split type, the ratio between the CFAL and AFCL methods would be extremely high if the number of series with high $P_{top\ 10\%}/P$ ratios was low.

## 7. CONCLUSIONS

All the $P_{top\ 10\%}$ and $P_{top\ 10\%}/P$ ratios reported by the National Science Board, European Commission, Leiden Ranking and others are real counts and thus correct. The issue raised by this study regards whether these indicators reveal the contribution of countries and institutions to the progress of knowledge, which is reported in publications that are 1000 or 10,000 times less frequent than the top 10% publications. Therefore, in formal terms, $P_{top\ 10\%}$ and the $P_{top\ 10\%}/P$ ratio are not indicators of the contribution to the progress of knowledge; they only allow for the calculation of real indicators. On the basis of this property, $P_{top\ 10\%}$ and the $P_{top\ 10\%}/P$ may be used as indicators of scientific advances. This study shows that the theoretical framework that supports this possibility is only correct when applied to homogeneous research groups. From a theoretical viewpoint, for countries and most institutions, this condition is not fulfilled, and $P_{top\ 10\%}$ and the $P_{top\ 10\%}/P$ are not indicators of the progress of knowledge.

Despite this conclusion, country and university rankings based on $P_{top\ 10\%}$ and $P_{top\ 10\%}/P$ ratios have been habitually taken as indicators of the contribution to the progress of knowledge without criticisms, except in the case of Japan. This strongly suggests that, among most countries and institutions, the distributions of the units of research are similar and the assumption that the $P_{top\ 10\%}$ and $P_{top\ 10\%}/P$ ratio are indicators of the progress of knowledge may be reasonable because they are similarly inaccurate; for this reason, this inaccuracy has gone unnoticed. However, evaluations of Japan reveal it.

Now that this inaccuracy has been discovered, a reconsideration of the previous assumption that $P_{top\ 10\%}$ (or $P_{top\ 1\%}$) is an indicator of the contribution to the advancement of knowledge and finding a procedure for the correct calculation of $P_{top\ 0.01\%}$ or a similar indicator become urgent goals. The identification of the units of research seems difficult, but either this identification or the formulation of new indicators is absolutely necessary. It is possible that there are no other countries for which deviations are as large as for Japan, but even smaller deviations may change the evaluation of countries and institutions substantially. Bringing these rankings closer to reality is currently a challenge in scientometrics.

**FUNDING INFORMAMATION**

This work was supported by the Spanish Ministerio de Ciencia e Innovación, Grand Number PID2020-113455GB-I00

**REFERENCES**

Abramo, G., & D'Angelo, C. A. (2016a). A farewell to the MNCS and like size-independent indicators: Rejoinder. *Journal of Informetrics*, *10*, 679-683.

Abramo, G., & D'Angelo, C. A. (2016b). A farewell to the MNCS and like size-indpendent indicators. *Journal of Informetrics*, *10*, 646-651.

Aksnes, D. W., Langfeldt, L., & Wouters, P. (2019). Citations, citation indicators, and research quality: An overview of basic concepts and theories. *SAGE Open*, *January-March: 1-17*.

Bornmann, L., Leydesdorff, L., & Mutz, R. (2013). The use of percentile rank classes in the analysis of bibliometric data: opportunities and limits. *Journal of Informetrics*, *7*, 158-165.

Bornmann, L., Ye, A., & Ye, F. (2018). Identifying landmark publications in the long run using field-normalized citation data. *Journal of Documentation*, *74*, 278-288.

Brito, R., & Rodríguez-Navarro, A. (2018). Research assessment by percentile-based double rank analysis. *Journal of Informetrics*, *12*, 315-329.

Cameron, W. B. (1963). *Informal Sociology, a casual introduction to sociological thinking* Randon House.

Charlton, B. G. (2007). Scientometric identification of elite 'revolutionary science' research institutions by analysis of trends in Nobel prizes 1947-2006. *Medical Hypotheses*, *68*, 931-934.

European Commission. (2018). *Science, Research and Innovation Performance of the EU 2018. Strengthening the foundations for Europe's future*. RTD-PUBLICATIONS. https://doi.org/10.2777/14136

European Commission. (2020). *Science, Research and Innovation Performance of the EU 2020. A fair, green and digital Europe*. Publication Office of the European Union.

Golosovsky, M. (2021). Universality of citation distributions: A new understanding. *Quantitative Science Studies*, *2*, 527-543.

Kuhn, T. (1970). *The structure of scientific revolutions*. University of Chicago Press.

Limpert, E., Stahel, W. A., & Abbt, M. (2001). Log-normal distributions across the sciences: Keys and clues. *BioScience*, *51*, 341-352.

Narin, F. (1976). *Evaluative bibliometrics: The Use of Publication and Citation Analysis in the Evaluation of Scientific Activity*. Computer Horizon Inc.

National Science Board. (2016). *Science and Engineering Indicators 2016*. National Science Fundation.

National Science Board. (2022). *Science and Enginering Indicators 2022: The State of U.S. Science and Engineering*. NSB-2022-1.

Pendlebury, D. A. (2020). When the data don't mean what they say: Japan's comparative underperformance in citation impact. In C. Daraio & W. Glanzel (Eds.), *Evaluative Informetrics: The Art of Metrics-based Research Assessment*. Spriger.

Radicchi, F., Fortunato, S., & Castellano, C. (2008). Universality of citation distributions: toward an objective measure of scientific impact. *Proc. Natl. Acad. Sci. USA*, *105*, 17268-17272.

Redner, S. (2005). Citation statistics from 110 years of *Physical Review*. *Physics Today*, *58*, 49-54.

REF2014 (2011). Assessment framework and guidance on submissions. https://http://www.ref.ac.uk/2014/pubs/2011-02/ accessed on October 2018

Rodríguez-Navarro, A. (2011). Measuring research excellence. Number of Nobel Prize achievements versus conventional bibliometric indicators. *Journal of Documentation*, *67*, 582-600.

Rodríguez-Navarro, A., & Brito, R. (2018a). Double rank analysis for research assessment. *Journal of Informetrics*, *12*, 31-41.

Rodríguez-Navarro, A., & Brito, R. (2018b). Technological research in the EU is less efficient than the world average. EU research policy risks Europeans' future. *Journal of Informetrics*, *12*, 718-731.

Rodríguez-Navarro, A., & Brito, R. (2019). Probability and expected frequency of breakthroughs – basis and use of a robust method of research assessment. *Scientometrics*, *119*, 213-235.

Rodríguez-Navarro, A., & Brito, R. (2020). Like-for-like bibliometric substitutes for peer review: advantages and limits of indicators calculated from the ep index. *Research Evaluation*, *29*, 215-230.

Rodríguez-Navarro, A., & Brito, R. (2021). Total number of papers and in a single percentile fully describes reserach impact-Revisiting concepts and applications. *Quantitative Science Studies*, *2*, 544-559.

Rodríguez-Navarro, A., & Brito, R. (2022a). The extreme upper tail of Japan's citation distribution reveals its research success. *arXiv:2201.04031*.

Rodríguez-Navarro, A., & Brito, R. (2022b). The link between countries' economic and scientific wealth has a complex dependence on technological activity and research policy. *Scientometrics*, *127*, 2871-2896.

Schlagberger, E. M., Bornmann, L., & Bauer, J. (2016). At what institutions did Nobel lauretae do their prize-winning work? An analysis of bibliographical information on Nobel laureates from 1994 to 2014. *Scientometrics*, *109*, 723-767.

Thelwall, M. (2016a). Are there too many articles? Zero inflated variants of the discretised lognormal and hooked power law. *Journal of Informetrics*, *10*, 622-633.

Thelwall, M. (2016b). Citation count distributions for large monodisciplinary journals. *Journal of Informetrics*, *10*, 863-874. https://doi.org/http://dx.doi.org/10.1016/j.joi.2016.07.006

Thelwall, M. (2016c). The precision of the aritmetic mean, geometric mean and percentiles for citation data: an experimental simulation modelling approach. *Journal of Informetrics*, *10*, 110-123.

Viiu, G.-A. (2018). The lognormal distribution explains the remarkable pattern documented by characteristic scores and scales in scientometrics. *Journal of Informetrics*, *12*, 401-415.

Waltman, L. (2016). A review of the literature on citation impact indicators. *Journal of Informetrics*, *10*, 365-391.

Waltman, L., & Schreiber, M. (2013). On the calculation of percentile-based bibliometric indicators. *Journal of the American Society for information Science and Technology*, *64*, 372-379.

Waltman, L., van Eck, N. J., & van Raan, A. F. J. (2012). Universality of citation distributions revisited. *Journal of the American Society for information Science*, *63*, 72-77.